# Recent Advances in Medical Image Classification

Loan Dao, Ngoc Quoc Ly

Dept. of Computer Vision and Cognitive Cybernetics_University of Science, VNUHCM, Ho Chi Minh, Vietnam, Viet Nam National University, Ho Chi Minh City, Vietnam

*Abstract*—Medical image classification is crucial for diagnosis and treatment, benefiting significantly from advancements in artificial intelligence. The paper reviews recent progress in the field, focusing on three levels of solutions: basic, specific, and applied. It highlights advances in traditional methods using deep learning models like Convolutional Neural Networks and Vision Transformers, as well as state-of-the-art approaches with Vision-Language Models. These models tackle the issue of limited labeled data, and enhance and explain predictive results through Explainable Artificial Intelligence.

## I. Introduction

Medical Image Classification (MIC), a crucial integration of Artificial Intelligence (AI) and Computer Vision (CV), is revolutionizing image-based disease diagnosis. By categorizing medical images into specific disease classes, MIC enhances diagnostic accuracy and efficiency. Utilizing various imaging modalities like X-rays, CT scans, MRI, and ultrasound, MIC systems cater to specific clinical needs. Incorporating state-of-the-art technologies, MIC optimizes classification accuracy, leading to precise diagnoses and improved patient care.

*1) The importance of MIC*: The ability to interpret medical images accurately and efficiently is crucial for timely and effective patient care. However, manual image analysis can be time-consuming and prone to human error. MIC, leveraging AI and CV, offers automated analysis and classification of medical images, leading to several benefits:

*a) Improved diagnostic accuracy*: MIC systems can detect subtle patterns and features at the pixel level that may be missed by human observers, leading to more accurate diagnoses.

*b) Reduced workload for physicians*: Automating image analysis frees up valuable time for physicians, allowing them to focus on patient interaction and complex decision-making.

*c) Enhanced efficiency*: MIC systems can process large volumes of images quickly, leading to faster diagnoses and treatment decisions.

*d) Improved patient outcomes*: Ultimately, the improved accuracy and efficiency of MIC contribute to better patient outcomes and overall healthcare quality.

*2) Challenges and the need for transparency*: While MIC offers immense potential, challenges remain. Hospital overload, physician burnout, and the risk of misdiagnosis necessitate robust and reliable MIC systems. Transparency and explainability are crucial for building trust among stakeholders. Explainable AI (XAI) addresses this need by providing insights into the decision-making process of MIC models, allowing physicians to understand the rationale behind classifications and make informed decisions.

*3) Advancements in MIC*: Recent advancements in MIC have significantly enhanced its capabilities. Large-scale Medical Vision-Language Models (Med-VLMs) trained on extensive datasets of image-caption pairs enable a deeper understanding of visual information, leading to more accurate and generalizable models. Additionally, novel network architectures like transformers and multi-task learning approaches have further improved performance and efficiency. Few-shot and zero-shot learning have also made significant contributions to MIC. Few-shot learning allows models to classify images with minimal labeled examples, beneficial in fields where obtaining large labeled datasets is challenging. Zero-shot learning enables models to classify images from unseen classes by leveraging knowledge transfer from related tasks. Combined with Explainable AI (XAI) techniques, these approaches not only explain results and increase model reliability but also optimize outcomes, enhancing system accuracy and performance. This comprehensive understanding and improved reliability facilitate their integration into clinical practice with high confidence and precision, ultimately leading to better patient outcomes and more efficient healthcare processes.

*4) Exploring MIC across three levels of solution*: To fully grasp the current state of MIC, this paper delves into three distinct levels:

*a) Level 1*: Basic Models: This level examines the fundamental theoretical models including MIC, including learning models, basic network architectures, and XAI techniques.

*b) Level 2*: Task-Specific Models: This level explores specific theoretical models and network architectures tailored to particular MIC tasks, such as single-task and multi-task classification.

*c) Level 3*: Applications: This level surveys prominent applications of MIC within the medical community, highlighting recent research trends and real-world implementations.

*5) Contributions and structure*: This article makes several key contributions:

*a) Comprehensive review*: It provides a thorough and systematic review of recent advancements in MIC, offering valuable insights for researchers and practitioners.

*b) Highlighting key developments*: It identifies and discusses significant breakthroughs, including VLMs, transformer-based architectures, multitask models, and progress in XAI, which not only explain prediction results but also enhance the performance of MIC. Notably, recent advancements in zero-shot learning and few-shot learning address data scarcity in the medical field and mitigate model overfitting.

*c) Addressing challenges and proposing solutions*: It explores challenges in MIC and proposes effective solutions to improve classification algorithms and systems.

*d) Exploring current issues*: It delves into pressing problems surrounding recent advancements in MIC, providing a deeper understanding of the evolving research landscape.

The remainder of the paper is structured as follows (Fig. 1): Section II overviews of recent advancements across three levels. Sections III to V detail each level. Section VI addresses challenges and proposes solutions. Section VII concludes and highlights future research directions. TABLE I. lists the abbreviations used.

By comprehensively exploring recent advancements in MIC, this article aims to contribute to the development of more effective and reliable classification systems, ultimately improving patient care and outcomes.

This comprehensive survey demonstrates the multi-faceted nature of medical image classification across various levels of solutions, providing researchers and practitioners with a holistic view of the field's current state and future directions. By synthesizing recent advancements in MIC across fundamental models, task-specific architectures, and real-world applications, this article not only addresses current challenges but also contributes significantly to the ongoing research in the field, offering valuable insights for future developments.

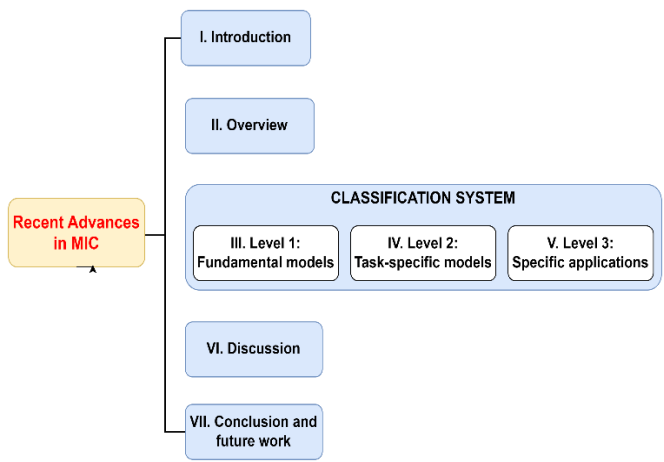

Fig. 1. Overview of paper organization.

TABLE I. LIST OF COMMON ABBREVIATIONS

| Abbreviation | Full Form |
|---|---|
| AI | Artificial Intelligence |
| CAD | Computer-Aided Diagnosis |
| CNN | Convolutional Neural Network |
| CV | Computer Vision |
| DL | Deep Learning |
| DNN | Deep Neural Network |
| FSL | Few-shot learning |
| Med-VLM | Medical Visual-Language Model |
| MIC | Medical Image Classification |
| MTL | Multitask Learning |
| ML | Machine Learning |
| NLP | Natural Language Processing |
| SOTA | State-of-the-Art |
| VLM | Vision-Language Model |
| XAI | eXplainable Artificial Intelligence |
| ZSL | Zero-shot learning |

## II. OVERVIEW OF RECENT ADVANCES IN MIC ACROSS THREE LEVELS OF CLASSIFICATION SYSTEMS

This section explores the evolving landscape of MIC through a standard three-level classification system framework. Each level serves a distinct purpose, building upon the foundations of the preceding one. TABLE II. provides a comprehensive overview of recent advancements in MIC across these three levels, highlighting their functionalities and advantages. This structured approach facilitates a deeper understanding of the current SOTA and the interconnected nature of progress within the field.

The proposed methods in this survey address key challenges in MIC:

- Med-VLMs leverage visual and textual data to mitigate limited labeled data issues, enhancing model robustness and generalizability.

- Few-shot and zero-shot learning techniques enable classification of rare conditions with minimal training examples.

- Transformer-based architectures and CNN hybrids capture both local and global features, improving complex medical image comprehension.

- XAI integration enhances interpretability, fostering trust and adoption in clinical settings.

These approaches represent targeted solutions to specific MIC challenges, demonstrating the field's adaptability to clinical needs. By addressing data scarcity, rare condition classification, feature extraction, and interpretability, these methods contribute to advancing AI-driven medical image analysis.

## III. Level 1 of MIC (Fundamental Models)

Level 1 includes learning models, fundamental network architectures and backbone DNN, and XAI. This level plays an essential role in developing systems at the subsequent levels.

### A. Learning Model

*1) Unimodal learning in MIC*: The evolution of learning models has significantly impacted the field of MIC, offering solutions to challenges like manual data labeling and limited generalization capacity. TABLE III. provides a concise comparison of various unimodal learning models commonly employed in MIC, highlighting their key characteristics and suitability for different scenarios. Selecting an optimal learning model for MIC tasks (see Table III) requires careful consideration of data availability, labeling costs, privacy requirements, and performance expectations. While supervised learning is powerful when labeled data is abundant, data annotation limitations and privacy concerns necessitate exploring alternative paradigms. Semi-supervised, weakly-supervised, active learning, meta-learning, federated learning, and self-supervised learning offer promising avenues to address these challenges, fostering the development of more efficient and generalizable MIC systems. Leveraging these diverse approaches allows researchers and practitioners can unlock the full potential of MIC, ultimately leading to improved patient care and clinical outcomes.

*2) Multimodal learning with med-VLMs in MIC*: Bridging the semantic gap between visual and textual information is crucial for effective MIC. VLMs integrate Computer Vision and Natural Language Processing, enabling a comprehensive understanding of medical data. This section explores the role of clinical and paraclinical data in Medical-VLMs (Med-VLMs) and surveys SOTA Med-VLMs for MIC.

*a) Clinical and paraclinical data in Med-VLMs*: To better understand the distinct roles and characteristics of clinical and paraclinical data within Med-VLMs, TABLE I. It provides a comparative analysis.

Clinical data provides valuable context for interpreting paraclinical images, while paraclinical data offers objective visualizations of potential abnormalities. Med-VLMs leverage both data types to enhance diagnostic accuracy and provide a holistic understanding of patient health.

*b) State-of-the-Art (SOTA) Med-VLMs in MIC*: Several advanced Med-VLMs have demonstrated remarkable performance in MIC tasks, utilizing sophisticated techniques such as transformer architectures, attention mechanisms, and pre-training on large datasets. TABLE V. summarizes SOTA Med-VLMs for MIC.

TABLE II. Overview of the Three-Level Solution Framework for Medical Image Classification

| Level | Content | Specific solutions | Explaination |
|---|---|---|---|
| 1 | Learning model | • **Unimodal learning:** Supervised learning, unsupervised learning, semi-supervised learning, weakly supervised learning, active learning, meta-learning, federated learning, self-supervised learning.<br>• **Med-VLMs:** BiomedCLIP [1], XrayGPT [2], M-FLAG [3], and MedBLIP [4].<br>• **Some remarkable methods:**<br>  ○ **Few-shot learning:** BioViL-T [5], PM2 [6], and DeViDe [7].<br>  ○ **Zero-shot learning:** MedCLIP [8], CheXZero [9], and MedKLIP [10]. | The evolution of learning models **from unimodal to multimodal**, exemplified by the emergence of **Med-VLM**, represents a significant advancement in the field. Few-shot and zero-shot learning models further enhance the ability to classify medical images with minimal or no labeled data, making them effective for rare and novel diseases. |
| 1 | Architectures of fundamental networks and backbone DNN | • **CNN**: VGGNet [11], GoogleNet [12], ResNet [13], and EfficienNet [14].<br>• **GNN**: Graph Convolution Networks (GCN) [15], and GAT [16].<br>• **Transformer**: ViT [17], DeiT [18], TransUnet [19], TransUnet+ [20], and TransUnet++ [21]. | Evolution of fundamental network architectures in image classification, including **CNNs, GNNs, and Vision Transformers**, as well as their respective backbone DNNs. |
| 1 | XAI | • **For CNN:** LIME [22], SHAP [23], CAM-based (CAM [24], GradCAM [25], and GradCAM++ [26]).<br>• **For Transformer:** ProtoPFormer[27], X-Pruner [28], and GradCam for Vision Transformer [29]. | XAI is applied for CNN Architecture and Vision Transformer |
| 2 | Specific DNN architectures and Med-VLM for single task (classification) | • **CNN:** Unet [30], Unet ++ [31], SNAPSHOT ENSEMBLE [32], and PTRN [33].<br>• **GNN:** CCF-GNN [34] and GazeGNN [35].<br>• **Transformer:** SEViT [36] and MedViT [37].<br>• **Med-VLM:** BERTHop [38], KAD [39], CLIPath [40], and ConVIRT [41]. | Specialized network architectures have achieved high performance in MIC.<br><br>Med-VLMs for MIC. |
| 2 | Specific DNN architectures and Med-VLM for multitask (classification and segmentation) | • **CNN:** Mask-RCNN-X101 [42] and Cerberus [43].<br>• **GNN:** MNC-Net [44] and AMTI-GCN [45].<br>• **Transformer:** TransMT-Net [46] and CNN-VisT-MLP-Mixer [47].<br>• **Med-VLM:** GLoRIA [48], ASG [49], MeDSLIP [50], SAT [51], CONCH [52], and ECAMP [53]. | MIC is advancing with multi-tasking. Classifying disease segments often excels over whole image analysis. The advent of Med-VLMs for multi-tasking enhances precision and depth of analyses. |
| 3 | Specific applications | • Breast Cancer [54] [55], tuberculosis [56], eye disease diagnosis [57][58], skin cancer diagnosis [59] [60], bone disease [61] - [63], other pathological [64] [65]<br>• Cancer, brain, tumor, lesion, lung, breast, eye, etc. | Surveying prominent applications significant to the medical community. Recent research trends in MIC (2020 - 2024) and cancer statistics for 2024 |

TABLE III. COMPARISON OF LEARNING MODELS IN MIC

| Learning model | Data Availability | Labeling cost | Operating principles | Balance | Applications |
|---|---|---|---|---|---|
| Supervised Learning | Labeled data required | High | Learns input-output mapping from labeled data | High performance with sufficient labeled data | Tumor detection, organ segmentation, classification |
| Unsupervised Learning | Unlabeled data only | Low | Finds patterns and structures in data without explicit supervision | Lower performance, useful for discovering underlying structures | Clustering similar images, anomaly detection |
| Semi-supervised Learning | Labeled and unlabeled data | Moderate | Utilizes a combination of labeled and unlabeled data to improve model performance | Higher performance than unsupervised learning with less labeling effort | Classification with limited labeled data |
| Weakly Supervised Learning | Weak supervision (coarse or image-level labels) | Moderate to low | Learns from partially labeled or noisy data | Scalability with reasonable performance | Image-level diagnosis tasks |
| Self-supervised Learning | Unlabeled data | Low | Generates supervisory signals from the input data itself | Balances model performance with labeling effort by leveraging unlabeled data | Efficient use of unlabeled data to pre-train models for downstream tasks |
| Active Learning | Small initial labeled dataset, actively selects informative samples | Initially high, decreases over time | Actively selects the most informative samples to be labeled | Balances model performance with labeling effort | Reducing labeling effort by prioritizing informative images |
| Meta-Learning | Diverse set of tasks for meta-training | High initially, potentially low for downstream tasks | Learns to learn from different tasks, improving adaptation to new tasks with limited data | Balances adaptation to new tasks with reduced need for extensive labeled data | Efficient adaptation to new imaging modalities or diseases |
| Federated Learning | Decentralized data across multiple devices/institutions | Varies depending on data distribution | Collaboratively trains a global model while keeping data localized | Balances model performance with data privacy and availability | Collaborative model training across institutions without sharing sensitive data |

To summary, Med-VLMs show significant potential for advancing MIC by effectively integrating clinical and paraclinical data. Key takeaways from the surveyed models include the effectiveness of transfer learning, model optimization techniques, integration of medical knowledge, and the development of multi-task models. These advancements pave the way for more accurate, efficient, and comprehensive diagnostic support tools in healthcare.

*3) Some remarkable methods*

*a) Few-shot learning in MIC*: In the medical imaging domain, few-shot learning (FSL) techniques are crucial due to the scarcity of labeled data and the dynamic nature of disease patterns. FSL enables accurate classification and diagnosis from a limited number of training samples, leveraging meta-learning and transfer-learning principles.

Core Principles:

Meta-learning: Models are trained on diverse medical imaging tasks to learn a shared representation that can be quickly adapted to new tasks with few examples, optimizing for rapid adaptation to new data.

Transfer learning: Pre-trained models on large medical datasets are fine-tuned on smaller, specific datasets to improve performance on the target task, such as disease classification or anomaly detection.

Relevant Med-VLM Models:

- BioViL-T [5] is a self-supervised learning approach that leverages temporal information within longitudinal medical reports and images to enhance performance on medical vision-language tasks. It utilizes a hybrid CNN-Transformer architecture for encoding visual data and a text model pretrained with contrastive and masked language modeling objectives. This approach enables BioViL-T to learn robust representations of medical concepts by capturing both visual and temporal relationships present in longitudinal data. The model's strength lies in its ability to transfer knowledge from diverse sources, leading to improved performance in few-shot settings.

- PM2 [6] introduces a novel multi-modal prompting paradigm for few-shot medical image classification. Its key strength lies in leveraging a pre-trained CLIP model and learnable prompt vectors to effectively bridge visual and textual modalities. This approach enables PM2 to achieve impressive performance in few-shot settings, surpassing existing methods on various medical image classification benchmarks.

- DeViDe [7] is a novel transformer-based approach that leverages open radiology image descriptions to align diverse medical knowledge sources, handling the complexity of associating images with multiple descriptions in multi-label scenarios. It guides medical image-language pretraining using structured medical knowledge, enabling more meaningful image and language representations for improved performance in downstream tasks like medical image classification and captioning.

Advantages:

- Data Efficiency: Reduces the need for large amounts of labeled data, making it feasible to develop models with limited resources.

- Flexibility: Can quickly adapt to new tasks with minimal data, which is crucial in dynamic environments like medical imaging.

Disadvantages:

- Performance: May be less effective compared to models trained on large, fully labelled datasets.
- Complexity: Requires careful design of task sets for training to ensure generalizability and robustness.

*b) Zero-shot learning in MIC*: Zero-shot learning (ZSL) enables the classification of unseen classes by leveraging semantic relationships between known and unknown classes. ZSL's core principle is to use auxiliary information, such as textual descriptions, to bridge the gap between seen and unseen classes, thereby expanding AI systems' diagnostic capabilities.

Core Principles:

- Semantic Embeddings: Align visual features with semantic representations (e.g., word embeddings) to infer the class of unseen instances by creating a shared space where both visual and semantic data coexist.
- Knowledge Transfer: Utilize knowledge from known classes to predict the properties of unknown classes based on their semantic descriptions, effectively transferring learned information across domains.

Common Models:

- MedCLIP [8] uses contrastive learning from unpaired medical image-text data to improve representation learning and zero-shot prediction, achieving strong performance even with limited data.
- CheXZero [9] is a deep learning model specifically for chest X-ray classification, utilizing pre-trained CNNs and fine-tuning on labelled data to achieve high accuracy in identifying thoracic diseases.
- MedKLIP [10] leverages medical knowledge during language-image pre-training in radiology, enhancing its ability to handle unseen diseases in zero-shot tasks and maintaining strong performance even after fine-tuning.

These models represent significant advancements in medical image classification, demonstrating impressive results and addressing the unique challenges posed by healthcare data.

Advantages:

- Scalability: Enables classification of novel classes without prior training examples, making it highly scalable and versatile.
- Flexibility: Expands the diagnostic capabilities of AI systems to include rare and novel diseases, which are often not well-represented in training datasets.

Disadvantages:

- Accuracy: Performance may be lower compared to models trained specifically on the classes of interest, particularly for highly dissimilar unseen classes.
- Dependency on Semantic Descriptions: Requires accurate and rich semantic information to function effectively, which can be a limitation if such data is not available.

Overall, few-shot and zero-shot learning models address the challenge of limited labeled data in medical image classification. FSL adapts quickly to new tasks with minimal training samples, while ZSL uses semantic relationships to diagnose rare and novel diseases. Each approach has unique advantages and limitations that must be considered when designing MIC systems. Understanding these principles is crucial for developing effective and reliable MIC models.

*B. Architectures of Fundamental Networks and Backbone DNN*

MIC has significantly shifted from traditional machine learning methods to deep learning approaches. This review focuses on fundamental DL architectures commonly used in MIC, including Convolutional Neural Networks (CNNs), Graph Neural Networks (GNNs), and Transformers. These architectures have shown remarkable efficacy in automatically learning hierarchical feature representations and achieving state-of-the-art performance in various MIC tasks.

*1) Convolutional Neural Networks (CNNs)*: CNNs have become the cornerstone of MIC due to their ability to automatically learn hierarchical feature representations. Inspired by the human visual cortex (Fig. 2 [66]), CNNs excel at capturing local features within images, making them ideal for tasks like disease detection, organ segmentation, and anomaly identification. This section explores the core components of CNNs and their contributions to feature extraction and classification, followed by a review of popular CNN architectures and their advancements in MIC.

*a) Core components of CNNs*: TABLE VI. summarizes the core components of a CNN and their functions in feature extraction and class prediction.

These components work synergistically to enable CNNs to learn intricate features from medical images, leading to accurate classification.

TABLE IV.    COMPARISON OF CLINICAL AND PARACLINICAL DATA

| Feature | Clinical Data | Paraclinical Data |
|---|---|---|
| Source | Direct interaction with healthcare professionals | Direct interaction with healthcare professionals |
| Nature | Text-based (medical history, symptoms, physical exam findings) | Image-based (internal body structures) |
| Role | Subjective assessment of patient condition | Objective visualization of abnormalities |
| Usage in VLMs | Provides context and complements image interpretation | Serves as primary input for image analysis and classification |

TABLE V.    PROMINENT MED-VLMS IN MIC

| Med-VLMs | Principle | Encoders and fusion method | Pre-trained objectives | Implementation Details | Performance Metrics | Key Contributions |
|---|---|---|---|---|---|---|
| BiomedCLIP [1] | Adapts CLIP for biomedical domains | Language encoder: PubMedBERT Vision encoder: ViT Fusion method: late fusion | Cross-modal global contrastive learning | Toilored batch size and patch dropout strategy for efficiency. | Pcam: 73.41, LC25000 (lung): 65.23, LC25000 (colon): 92.98, TCGA-TIL: 67.04, RSNA: 78.95 | Superior zero-shot and few-shot classification. Outperfrms SOTA models on diverse iomedical dataset, robust image encoder. |
| XrayGPT [2] | Summarizes chest X-rays by aligning MedClip with Vicuna. | Language encoder: Vicuma Vision encoder: MedCLIP Fusion method: early fusion | Hybrid: Image-report matching and mixed objectives | Fine-tuned Vicuna on curated reports | Interactive summaries from radiology reports | Integration of medical knowledge through interactive summaries, enhancing the interpretability and usability of diagnostic results. |
| M-FLAG [3] | Frozen language model, orthogonality loss for harmonized latent space. | Language encoder: CXR-BERT (frozen) Vision encoder: ResNet50 Fusion method: late fusion | Hybrid: Image-text contrastive learning and language generative | Potential for classification, segmentation, object detection | Outperforms existing MedVLP approaches, 78% parameter reduction | Model optimization and efficiency, achieving high performance with reduced parameters. |
| MedBLIP [4] | Bootstraps VLP from 3D medical images and texts | Language encoder: BioMedLM Vision encoder: ViT-G14 (EVA-CLIP) Fusion method: late fusion | Global and local contrastive learning | Combines pre-trained vision and language models | SOTA zero-shot classification of Alzheimer's disease | Efficient 3D medical image processing facilitates classifying complex conditions with minimal labeled data. |

*b) Popular CNN architectures*: A Historical Perspective: The evolution of CNN architectures has been driven by continuous innovation in addressing challenges and improving performance. TABLE VII. highlights key milestones:

CNN architectures offer unique advantages and have demonstrably excelled in image classification tasks. Their capacity to learn intricate features and generalize to new data underscores their value in advancing image analysis and related research fields. Ongoing research promises further innovations in CNN architecture and training methodologies, leading to increasingly accurate and efficient image classification systems. This progress holds particular significance for the medical domain, where precise image classification can directly impact diagnosis and patient care.

*2) Graph Neural Networks (GNNs)*: leveraging relationships in image data

GNNs offer a unique approach to image classification by representing images as graphs and exploiting the relationships between pixels or image regions. This allows GNNs to capture contextual information and learn more robust representations compared to traditional CNNs.

*a) GNN variants and their advantages*: Two prominent Graph Neural Network (GNN) variants demonstrate considerable potential in image classification: Graph Convolutional Networks (GCNs) and Graph Attention Networks (GATs).

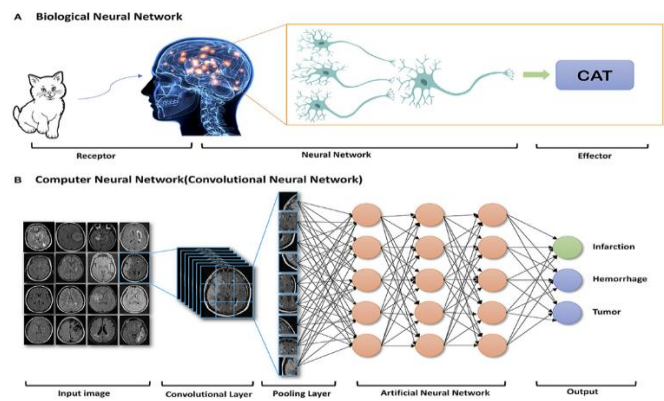

Fig. 2.    Illustration of convolutional neural networks (CNNs) inspired by biological visual mechanisms [66].

*(IJACSA) International Journal of Advanced Computer Science and Applications,*
*Vol. 15, No. 7, 2024*

TABLE VI. CNN COMPONENTS AND THEIR ROLES IN MIC

| Component | Function | Role in MIC |
|---|---|---|
| Convolutional Layer | Applies filters to extract local features (edges, textures) | Hierarchical feature extraction, capturing spatial relationships. |
| Activation Function (e.g., ReLU) | Introduces non-linearity for learning complex patterns. | Enables complex decision boundaries for accurate classification. |
| Pooling Layer (e.g., Max Pooling) | Down-samples feature maps to reduce dimensionality and improve invariance. | Improves robustness to image variations and reduces computational cost. |
| Fully-Connected Layer | Integrates local features into global patterns for image understanding. | Combines learned features for final class prediction. |
| Softmax Layer | Converts outputs into probability distribution over predicted classes. | Provides class probabilities for determining the most likely class. |

TABLE VII. POPULAR CNN ARCHITECTURES AND THEIR ADVANCEMENTS

| Architecture (Year) | Advancement | Key Technique |
|---|---|---|
| VGGNet (2014, [11]) | Achieved SOTA performance with increased depth | Small 3x3 filters for deeper networks |
| GoogleNet (2014, [12]) | Further reduced error rates with efficient architecture | Inception modules, 1x1 convolutions, global average pooling |
| ResNet (2015, [13]) | Enabled training of very deep networks | Residual blocks with skip connections to address vanishing gradients |
| EfficientNet (2019, [14]) | SOTA accuracy with fewer parameters | Compound scaling for optimal efficiency and performance |

- GCNs [15], by generalizing the convolution operation to graph data, effectively capture the local graph structure and relationships between nodes. This capability allows GCNs to leverage the inherent structural information within images for improved classification.

- GATs [16], on the other hand, introduce an attention mechanism to GNNs. This mechanism enables GATs to focus on relevant features within the graph, leading to improved feature extraction and ultimately, enhanced prediction accuracy. By selectively attending to important features, GATs can make more informed decisions during image classification.

*b) Benefits of GNNs for Image Classification*

- Modeling complex relationships: GNNs excel at capturing intricate dependencies between image elements, leading to better understanding of image context.

- Improved feature extraction: By considering relationships between nodes, GNNs can extract more informative and discriminative features for classification.

- Enhanced robustness: GNNs are less susceptible to noise and variations in image data due to their focus on structural information.

*c) Summary*: GNNs offer a valuable complementary approach to CNNs for image classification, particularly when dealing with data where relationships between elements are crucial. Their ability to leverage graph structures and learn contextual representations opens new avenues for improving accuracy and robustness in image classification tasks.

*3) Transformers*: Expanding Horizons in Image Classification

Transformers, initially designed for NLP, have emerged as powerful contenders in image classification. Unlike CNNs, transformers leverage self-attention mechanisms to capture global context and long-range dependencies within images, leading to richer feature representations.

*a) Contributions of transformers to image classification*

- Feature extraction: Vision transformers (ViTs [17]) split images into patches, embed them into vectors, and incorporate positional information. Self-attention mechanisms then assess the importance of each patch in relation to others, enabling the capture of global context and intricate features.

- Class prediction: A classification head on top of the final transformer encoder layer predicts the image class based on the learned global context. Parallel processing of patches enhances computational efficiency compared to sequential CNNs.

*b) Evolution of transformer architectures*

- Vision Transformer (ViT [17]): Introduced the transformer architecture to image classification, achieving impressive performance with patch-based processing and self-attention.

- Data-efficient image Transformers (DeiT [18]): Improved efficiency through knowledge distillation and efficient training strategies, achieving comparable results with fewer resources.

- Specialized variants (e.g., TransUnet [19], TransUnet+ [20], and TransUnet++ [21]): Combine transformers with U-Net architectures for enhanced feature extraction and accurate segmentation in medical imaging tasks.

*c) Addressing challenges:* Techniques like dropout, regularization, and efficient optimization algorithms mitigate overfitting and manage computational complexity in transformers.

In summation, the choice of architecture depends on the specific task and dataset characteristics. CNNs excel at local feature extraction, GNNs leverage relationships within data, and transformers capture global context and long-range dependencies. Understanding these strengths and weaknesses empowers researchers to select the most appropriate architecture for their MIC tasks.

*C. Explainable Artificial Intelligence (XAI) in MIC*

XAI techniques are crucial for fostering trust and understanding in MIC systems. Despite achieving human-level accuracy, the integration of automated MIC into clinical practice has been limited due to the lack of explanations for algorithmic decisions. XAI methodologies provide insights into the rationale behind the classification results of DL models, such as CNNs and Transformers, used in MIC tasks. By addressing the '**how**' and '**why**' behind predictive outcomes, XAI enhances the transparency and interpretability of MIC systems, contributing to their improved performance and acceptance in clinical settings.

*1) XAI methods in CNNs and transformers*: The field of XAI has witnessed significant advancements, particularly in the domain of MIC. This progress is evident in the evolution of XAI methods, transitioning from those primarily designed for CNNs to novel techniques tailored for Transformer architectures. The following tables provide a comparative analysis of recent advancements in XAI methods applied to CNNs (TABLE VIII. ), Transformers (TABLE IX. ) within the MIC domain, along with techniques used to enhance system performance (TABLE X. ).

TABLE VIII. XAI METHODS FOR CNNS IN MIC

| Method | Principle | How/Why Explanation | Methodological Approach | System Performance Impact |
|---|---|---|---|---|
| LIME [22] | Approximates complex models with simpler interpretable ones (e.g., linear regression) | Explains "how" by analyzing feature perturbation impact | Fits a simpler interpretable model to perturbed samples around an instance | Enhances local interpretability but may not capture global model behavior |
| SHAP [23] | Assigns feature contributions based on game theory principles. | Explains "how" and "why" by quantifying feature importance and interactions | Computes average feature contribution across all possible feature subsets | Provides global and local explanations, valuable for understanding complex models |
| CAM [24] | Visualizes image regions contributing most to a specific class. | Explains "why" by highlighting relevant regions | Combines feature maps and gradients to create a saliency map | Helps localize important features but lacks fine-grained details |
| Grad-CAM [25] | Improves CAM by incorporating gradient weights. | Explains "why" through visualization and "how" through contribution values on saliency maps | Computes gradients of class score with respect to feature maps for saliency map creation | Offers better localization and is widely used |
| Grad-CAM++ [26] | Refines Grad-CAM by addressing negative values and weight stability. | Explains "why" by enhancing visualization quality and "how" through weighted combination of positive and negative partial derivatives | Introduces Shapley values to estimate pixel contributions | Provides improved visual explanations and robustness |

TABLE IX. XAI METHODS FOR TRANSFORMERS IN MIC

| Method | Principle | How/Why Explanation | Methodological Approach | System Performance Impact |
|---|---|---|---|---|
| ProtoPFormer [27] | Interpretable image recognition using global and local prototypes | Explains "how" by utilizing prototypes to capture target features and "why" by addressing the need for improved interpretability in ViTs | Employs prototype-based XAI technique to enhance ViT interpretability | Achieves superior performance and visualization results compared to SOTA baselines |
| X-Pruner [28] | Explainable pruning framework for ViTs | Explains "how" by measuring each unit's contribution to class prediction using an explainability-aware mask | Adaptively searches layer-wise threshold based on explainability-aware mask values | Outperforms SOTA black-box methods with reduced computational costs and slight performance degradation |
| Grad-CAM for ViTs [29] | Visualization of VT decision-making | Explains "why" by revealing focus areas during ViT decision-making | Generates class activation maps for ViTs | Can enhance ViT model fine-tuning but requires further improvement |

TABLE X. XAI TECHNIQUES FOR ENHANCING SYSTEM PERFORMANCE

| Technique | Description | Impact on System Performance |
|---|---|---|
| Explainable Pruning | Pruning techniques like **X-Pruner** that utilize XAI to guide the **removal of less important** model components. | Reduces computational cost and model complexity while maintaining or improving performance. |
| Attention Visualization | Visualizing attention mechanisms in Transformers to **understand which parts of the input the model focuses on**. | Provides insights for **model improvement** and **debugging.** |
| Feature Importance Analysis | Techniques like **SHAP** that quantify **the importance of individual features** for model predictions. | Helps identify **key features and potential biases,** leading to improved model design and feature engineering. |
| Adversarial Training | Training models with adversarial examples to **improve robustness and generalizability**. XAI methods can be used to analyze the impact of adversarial attacks and guide the development of defense strategies. | Enhances model robustness and performance against adversarial attacks. |

*2) Discussion*: The tables above illustrate the diverse range of XAI methods available for both CNNs and Transformers in MIC. While CNN-based methods like LIME, SHAP, and Grad-CAM variants have been widely explored, the emergence of Transformers has led to the development of novel techniques like ProtoPFormer and X-Pruner. These methods offer unique advantages in terms of interpretability and performance improvement.

*a) Key Observations*

- Focus on Visual Explanations: Many XAI methods, particularly those applied to CNNs, emphasize visual explanations through saliency maps and other visualization techniques. This is crucial in MIC, where understanding the model's focus on specific image regions is essential for building trust and ensuring reliable diagnoses.

- Evolution from Local to Global Explanations: XAI methods have progressed from providing local explanations for individual predictions (e.g., LIME) to offering global interpretations of model behavior (e.g., SHAP). This allows for a more comprehensive understanding of the decision-making process.

- Integration with Model Optimization: Techniques like X-Pruner demonstrate the potential of integrating XAI with model optimization strategies like pruning. This allows for the development of more efficient and interpretable models.

*b) Future directions*

- Developing XAI methods specifically tailored for Transformer architectures: While existing techniques like Grad-CAM have been adapted for ViTs, further research is needed to explore methods that fully leverage the unique characteristics of Transformers.

- Combining XAI with other AI advancements: Integrating XAI with areas like federated learning and continual learning can lead to more robust and adaptable medical image classification systems.

- Standardization and Benchmarking: Establishing standardized evaluation metrics and benchmarks for XAI methods will facilitate fair comparisons and accelerate progress in the field.

*c) Enhancing performance and accuracy in MIC with XAI:* XAI techniques significantly improve the performance and accuracy of MIC systems by providing transparency and facilitating error detection and correction. These techniques help identify and rectify model shortcomings, leading to more reliable and effective MIC systems.

CNN-based XAI Techniques:

- LIME create interpretable models for individual predictions, helping to identify and correct misclassifications by highlighting important features.

- SHAP provide a unified measure of feature importance, allowing for precise identification of influential features and potential sources of errors.

- CAM-based Methods: These methods generate visual explanations by highlighting regions in the input image that influence the model's predictions, making it easier to spot and address inaccuracies.

Transformer-based XAI Techniques:

- ProtoPFormer uses prototypical parts to explain predictions, aiding in the identification of errors by comparing new instances with learned prototypes.

- X-Pruner prunes less important parts of the model, enhancing interpretability and helping to pinpoint and fix model weaknesses.

- GradCam for Vision Transformer adapts GradCAM for transformers, providing visual explanations that help in diagnosing and correcting errors in transformer-based MIC models.

Impact on MIC:

- Error Detection: XAI techniques make it easier to identify misclassifications and understand why they occur, enabling targeted corrections.

- Model Improvement: By revealing which features and regions are most influential, XAI helps refine model training and architecture, leading to better performance.

- Trust and Reliability: Enhanced transparency builds trust among clinicians, ensuring that MIC systems are more likely to be adopted and relied upon in clinical settings.

Some recent XAI techniques:

Recent studies have shown that using XAI methods such as Integrated Gradients can significantly enhance the performance of classification systems.

- A notable study by Apicella et al. (2023, [67]) investigated the application of Integrated Gradients, a technique from XAI, to enhance the performance of classification models. The study focused on three distinct datasets: Fashion-MNIST, CIFAR10, and STL10. Integrated Gradients were employed to identify and quantify the importance of input features contributing to the model's predictions. By analyzing these feature attributions, the researchers were able to pinpoint which features had the most significant impact on the model's output. The insights gained from Integrated Gradients were then used to refine the model. This involved adjusting the model parameters and structure to better capture the critical features identified by the XAI method. The study demonstrated that through this process of feature importance analysis and subsequent model optimization, the classification performance improved significantly across all tested datasets. This approach not only enhanced accuracy but also provided a deeper understanding of the model's decision-making process.

- Additionally, another study by Apicella et al. (2023, [68]) introduced an innovative method that also leveraged Integrated Gradients to boost classification system performance. This study proposed a soft masking scheme, wherein the explanations generated by Integrated Gradients were used to create masks that highlight important features while downplaying less relevant ones. The soft masking approach involved applying these masks during the training phase of the machine learning model. By focusing the model's attention on the most influential features as determined by Integrated Gradients, the training process became more efficient and effective. The experimental results from this study showed a marked improvement in model accuracy across the same datasets: Fashion-MNIST, CIFAR10, and STL10. The use of soft masks helped in reducing noise and enhancing the signal of critical features, thereby leading to better generalization and performance of the classification system.

All in all: XAI explanations enhance both models understanding and classification performance.

Summary of Level 1 Findings:

- Learning models have evolved from traditional supervised learning to advanced techniques like Med-VLMs, few-shot, and zero-shot learning, addressing data scarcity in medical imaging.
- Network architectures have progressed from CNNs to Transformers, with hybrid models showing promise in capturing both local and global features.
- XAI methods have become crucial for enhancing model interpretability and trust in clinical settings, with techniques like Grad-CAM and SHAP leading the way.
- The integration of these advancements has led to more robust, efficient, and interpretable models for medical image classification

## IV. LEVEL 2 OF MIC (TASK-SPECIFIC MODELS)

Expanding on initial network architectures, the second level focuses on specialized architectures for MIC. It takes a comprehensive approach, combining classification with segmentation through multitask learning models. This broad view deepens understanding of MIC network architectures, paving the way for specific applications.

*A. Recent Advances in Level 2 for Single Task*

*1) Specific DNN architectures for single task (classification)*: This review assesses recent advancements in DNN architectures for single-task classification in medical image analysis. It evaluates specialized architectures across CNNs, GNNs, and Transformers, considering methodology, datasets, effectiveness, advantages, and limitations. The comparative analysis, summarized in Table XITABLE XI. , highlights key developments and their implications for MIC, offering a comprehensive overview of the current state-of-the-art in the field.

Key insights:

*a) Adaptability and efficiency*: Unet and Unet++ demonstrate adaptability to new tasks and improved segmentation accuracy, though at the cost of increased parameters.

*b) Innovative approaches*: Snapshot Ensemble and GazeGNN introduce novel methods like GradCAM and eye-gaze data utilization, showcasing the potential of combining different data types and analytical techniques.

*c) Challenges in complexity and data requirements*: While architectures like PTRN and CCF-GNN show promise in specific tasks, they highlight the ongoing challenges of computational demands and the need for extensive training data.

*d) Future directions*: The evolution from CNN-based architectures to incorporating GNN and Transformer models indicates a shift towards more complex, yet potentially more effective methods for medical image classification. However, issues such as interpretability, computational efficiency, and data availability remain critical areas for future research.

This summary underscores the dynamic nature of deep learning research in medical image classification, emphasizing the need for continued innovation and exploration of new methodologies.

*2) Med-VLMs for MIC*: The recent rise of Med-VLMs has greatly influenced MIC. These models utilize NLP and CV to analyze medical images and text reports, enhancing diagnostic accuracy and efficiency. Table XII summarizes key Med-VLMs in MIC, highlighting their performance in zero-shot and few-shot learning scenarios:

Advancements and Impact

Med-VLMs demonstrate remarkable progress in MIC, particularly in scenarios with limited labeled data.

*a) Zero-shot learning*: Models like KAD showcase the ability to classify images of unseen pathologies without explicit training, highlighting the potential for real-world clinical applications.

*b) Few-shot learning*: CLIPath and ConVIRT achieve SOTA performance with minimal labeled data, reducing the burden of data annotation in clinical settings.

Future Directions

The field of Med-VLMs is rapidly evolving, with ongoing research exploring:

*a) Multi-modal learning*: Integrating diverse data modalities (e.g., images, text, genomics) for a more comprehensive understanding of diseases.

*b) Explainability and interpretability*: Enhancing transparency and trust in model predictions.

*c) Domain adaptation*: Adapting models to diverse clinical settings and populations.

Summary

Med-VLMs revolutionize MIC, promising improved diagnosis, treatment planning, and patient care. With ongoing research, they have the potential to transform healthcare, enabling more accurate, efficient, and personalized medicine.

*B. Recent Advances in Level 2 for Multitask (Classification and Segmentation)*

Multitask learning (MTL) is vital in MIC tasks, overcoming individual model limitations and boosting overall performance. A recent comprehensive study highlighted the substantial progress made in medical image segmentation using DNNs, leading to more accurate and efficient diagnostic processes [69]. By optimizing image segmentation and classification together, MTL provides numerous advantages:

*a) Mitigating data scarcity*: MTL leverages knowledge transfer across related tasks, enabling models to learn from complementary data sources and improve performance on the target task, even with limited data availability.

*b) Optimizing resource utilization*: By sharing feature representations across tasks, MTL optimizes the use of computational resources, leading to more efficient model architectures and reduced computational overhead.

*c) Learning robust shared representations*: MTL encourages the learning of shared features that are beneficial for both segmentation and classification tasks. These shared representations capture task-agnostic information, leading to improved generalization and performance across multiple MIC tasks.

MTL in MIC tasks effectively tackles challenges like data scarcity, resource constraints, and the necessity for robust, generalizable models. By leveraging synergies between related tasks, MTL enhances MIC systems' performance and efficiency, leading to better clinical decision-making and patient outcomes.

*3) Typical architecture for multitasking in MIC*: Researchers have investigated different MTL configurations like feature extraction, fine-tuning, and hybrids to match diverse medical imaging contexts and data availability. TABLE XIII. Surveys the latest notable DNN architectures using multitasking to boost MIC performance.

In summary, these MTL-based architectures demonstrate significant advancements in addressing data scarcity, improving resource efficiency, and leveraging shared representations to enhance medical image classification performance across various modalities and disease domains.

TABLE XI. SUMMARY OF KEY ARCHITECTURES FOR SINGLE TASK (MIC)

| Works | Method | Data | Effectiveness | Advantanges | Limitations |
|---|---|---|---|---|---|
| Unet (2015, [30]) | Supervised learning with encoder-decoder architecture | PhC-U373: 30 images DIC-HeLa: 35 images | High IOU scores (92% for PhC-U373, 77.5% for DIC-HeLa) | Accurate segmentation with limited data; adaptable to new tasks | Limited in extracting long-range information; lacks explanation for predictions |
| Unet++ (2018, [31]) | Supervised learning with redesigned skip paths | Cell nuclei, colon polyp, liver, lung nodule images | Cell nulei: 92.63, colon polyp: 33.45, liver: 82.90, lung nodule: 77.21 Improved IoU over Unet | Reduces semantic gap; improves accuracy and speed | Increases parameter count; lacks explanation for predictions |
| Snapshot Ensemble (2021, [32]) | Supervised learning with EfficientNet-B0, GradCAM | Malaria Dataset: 27558 erythrocyte images with equal cases of parasitized and uninfected cells. Source: CMC hospital in Bangladesh | High F1 score (99.37%) and AUC (99.57%) | Timely and accurate malaria diagnosis; uses GradCAM for explanations | Focused only on P. falciparum; not other species |
| PTRN (2022, [33]) | Supervised learning with DenseNet-201 | CheXpert: 224,316 digital CXRs; CheXphoto: 10,507 CXR | CheXpert: 0.896 CheXphoto-Monitor: 0.880 CheXphoto-Film: 0.802 meanAUC: 0.850 | Reduces cost of collecting natural data; eliminates negative impacts of projective transformation | Higher computation costs; untuned hyperparameters |
| CCF-GNN (2023, [34]) | Supervised learning with GNN | TCGA-GBMLGG, BRACS, Bladder Cancer, ExtCRC images | High AUC (0.912 for TCGA-GBMLGG) and accuracy | Effectively analyzes pathology images; represents cancer-relevant cell communities | Requires extensive training data; longer processing time |
| GazeGNN (2023, [35]) | Supervised learning with GNN | Chest X-ray: 1083 images | High accuracy (0.832) and AUC (0.923) | Captures complex relationships via graph learning without pre-generated VAMs | Needs eye-tracking devices for gaze data collection |
| SEViT (2022, [36]) | Supervised learning with Transformer | Chest X-ray: 7000 chest X-ray images (Normal or Tuberculosis) Fundoscopy (APTOS2019): diabetic retinopathy (DR) 3662 retina images (5 classes) | High accuracy (94.64% for Chest X-ray) and AUC | Detects adversarial samples by assessing prediction consistency | Full white-box settings not evaluated in natural image contexts |
| MedViT (2023, [37]) | Supervised learning with hybrid CNN-Transformer. | MedMNIST-2D: 12 biomedical datasets (CT, X-ray, Ultrasound, and OCT images) | Average accuracy of 0.851 and AUC of 0.942 | Reduces computational complexity; high generalization ability | Lacks precise hyperparameter tuning; employs two CNNs. |

TABLE XII. PERFORMANCE OF MED-VLMS IN MIC

| Model | Modality | Zero-shot Learning | Few-Shot Learning | Encoders and fusion method | Pre-trained objective | Key Features |
|---|---|---|---|---|---|---|
| BERTHop [38] | Chest X-ray | AUC: 98.12% | None | Language encoder: BlueBERT; Vision encoder: PixelHop++; Fusion method: Early fusion | Hybrid: matching and masking (masked language modelling) | Combines PixelHop++ and BlueBERT for effective visual-language fusion. |
| KAD [39] | Chest X-ray | Outperforms expert radiologists on multiple pathologies | Excels with few-shot annotations | Language encoder: PubMedBERT; Vision encoder: ResNet-50, ViT-16; Fusion method: Late fusion | Cross-modal global contrastive learning and hybrid with additional classification objective | Leverages medical knowledge graphs for improved zero-shot performance and auto-diagnosis. |
| CLIPath [40] | Pathology | Strong transferability | Efficient adaptation with limited data | Language encoder: BERT; Vision encoder: ResNet-50 or ViT; Fusion method: Early fusion | Contrastive learning | Fine-tunes CLIP using Residual Feature Connection for pathology image classification |
| ConVIRT [41] | Chest X-ray | Competitive performance | SOTA with few-shot annotations | Language encoder: BERT; Vision encoder: ResNet50; Fusion method: No fusion | Global contrastive learning | SOTA with few-shot annotations |

TABLE XIII. SUMMARY OF KEY ARCHITECTURES FOR MULTITASK LEARNING (MIC)

| Works | Method | Data | Effectiveness | Advantages | Limitations |
|---|---|---|---|---|---|
| Mask-RCNN-X101 (2021, [42]) | Supervised learning, Mask-RCNN-X101 architecture | 934 radiographs (667 benign, 267 malignant bone tumors) | Classification of bone tumors: 80.2% accuracy, 62.9% sensitivity, and 88.2% specificity. Bounding box placements: IoU of 0.52. Segmentation: mean Dice score 0.60. | Assists in diagnostic workflow by accurately placing bounding boxes, segmenting, and classifying primary bone tumors | Selection bias, inability to predict other diseases, fixed image resolution, lack of bone metastases and density information |
| Cerberus (2023, [43]) | Supervised learning, shared encoder (ResNet34) and independent decoders (U-Net) | Gland: (1602 GlaS + 3209 CRAG + 46346 generated), Lumen: 56358, Nuclei: 495179 | Segmentation: Nuclei 0.774, 0.560; Gland 0.908, 0.640; Lumen 0.666, 0.525; Classification: mAP 0.948, mF1 0.883 | Simultaneously predicts multiple tasks without compromising performance, publishes processed TCGA dataset | Performance enhancement in new tasks yet to be explored |
| MNC-Net (2023, [44]) | Supervised learning, graph encoder and cluster-layer | Parkinson's Progression Markers Initiative (PPMI) MRI data | ACC 95.50%, F1 95.49%, Prec 97.00%, Rec 94.42% | Early diagnosis of Parkinson's disease using clinical scores and brain regions, manages brain network complexity effectively | Limited to node-level tasks, does not capture all Parkinson's-related information |
| AMTI-GCN (2024, [45]) | Supervised learning, interpretation, feature sharing, and task-specific modules | AD-NC, AD-MCI, NC-MCI, MCIn-MCIp (186-393 samples) | NC-MCI: ACC 70.1, SEN 69.3, SPE 70.8, AUC 70.6, ADAS-Cog CC 0.477, MMSE CC 0.498; MCIn-MCIp: ACC 71.9, SEN 73.2, SPE 71.1, AUC 72.5, ADAS-Cog CC 0.485, MMSE CC 0.522 | Addresses limitations in binary Alzheimer's diagnosis and ignores task correlation | Did not explore potential correlations between ADAS-Cog, MMSE, and other factors like education level |
| TransMT-Net (2023, [46]) | Active learning, hybrid CNN-Transformer architecture | Polyp: 1,645 images | Seg.: DSC 77.76%, IoU 67.40%, 95% HD 21.62 mm; Class: Acc 96.94%, Pre 96.56%, Rec 96.52%, F1 96.54%; | Effectively addresses lesion classification and segmentation in GI tract endoscopic images | Slightly higher computational complexity, inferior segmentation performance with 70% training set, varied processing speed |
| CNN-ViST-MLP-Mixer (2024, [47]) | Supervised learning, hybrid CNN-ViT architecture and MLP-Mixer | BUSI: 789, UDIAT: 163 | Seg: BUSI (Acc 94.04, DC 83.42, IoU 72.56, Sen 80.10); UDIAT (Acc 97.88, DC 81.52, IoU 70.32, Sen 90.32); Class: Acc 86.00, Prec 86.11, Rec 86.02, F1 85.93, Sen 89.42, Spec 85.26 | Effectively captures local and high-level features in breast ultrasound images, enhances feature integration | Inability to monitor tumor's surrounding environment during diagnosis |

*4) Med-VLMs for multitask (classification and segmentation)*: Recent advancements in Med-VLMs have significantly improved the accuracy and efficiency of MIC by leveraging the power of multimodal AI. These models excel at handling multitask challenges, such as simultaneous classification and segmentation, leading to a more comprehensive understanding of medical images. Table XIV summarizes key Med-VLMs and their contributions to MIC.

These Med-VLMs demonstrate several key advancements in MIC:

*a) Enhanced medical knowledge*: Models like MedKLIP incorporate medical knowledge bases and text extraction techniques to improve understanding of medical images.

*b) Improved representation learning*: Techniques like attention mechanisms and contrastive learning enable models like GLoRIA and CONCH to learn more robust and efficient representations of medical images.

*c) Anatomical structure guidance*: ASG (IRA) and MeDSLIP leverage anatomical information to improve interpretability and clinical relevance, leading to more accurate classifications.

*d) Multitask capabilities*: Many of these models excel at both classification and segmentation tasks, providing a more comprehensive analysis of medical images.

*e) Zero-Shot and few-shot learning*: Several models, including GLoRIA and SAT, demonstrate strong performance even with limited labeled data, making them valuable in scenarios with scarce data resources.

Significantly, Med-VLMs are revolutionizing MIC by leveraging the power of multimodal AI and multitask learning. These models offer enhanced diagnostic precision, efficiency, and interpretability, ultimately leading to improved patient care and outcomes. As research in this area continues, we can expect even more powerful and versatile Med-VLMs to emerge, further transforming the field of medical imaging and healthcare as a whole.

TABLE XIV. COMPARISON OF MED-VLMS FOR MULTITASK MEDICAL IMAGE ANALYSIS

| Model | Encoders and fusion method | Pre-trained objective | Key innovations | Strengths |
|---|---|---|---|---|
| GLoRIA [48] | Language encoder: BioClinicalBERT Vision encoder: ResNet-50 Fusion method: late fusion | Global and local contrastive learning | Multimodal global-local approach, attention-weighted image regions | Data efficiency, zero-shot capabilities, excels in limited-label settings |
| ASG (IRA) [49] | Language encoder: BioClinicalBERT Vision encoder: ResNet-50 and ViT-B/16 Fusion method: late fusion | Contrastive learning and image tag recognition | Anatomical structure guidance, image-report alignment | Improved interpretability and clinical relevance, enhanced representation learning |
| MeDSLIP [50] | Language encoder: BioClinicalBERT Vision encoder: ResNet-50 Fusion method: late fusion | Hybrid: Prototypical contrastive learning and intra-image contrastive learning | Dual-stream architecture for disentangling anatomical and pathological information | Precise vision-language associations, improved performance in medical image captioning and report generation |
| SAT [51] | Language encoder: BioClinicalBERT Vision encoder: ResNet-50 Fusion method: late fusion | Contrastive learning | Semantic-aware transformer for integrating semantic information | Effective representation learning, excels in data/no-data recognition tasks |
| CONCH [52] | Language encoder: GPT-style Transformer Vision encoder: ViT-Base Fusion method: early fusion | Hybrid: Contrastive learning and captioning objective | Contrastive learning from captions for histopathology images | SOTA performance in histology image classification, segmentation, and retrieval tasks |
| ECAMP [53] | Language encoder: BERT Vision encoder: ViT-B/16 Fusion method: early fusion (multi-scale context fusion) | Hybrid: masked image modeling, masked language modeling, and context-guided super-resolution | Entity-centered context-aware pre-training, multi-scale context fusion | Enhanced text-image interplay, improved performance in downstream medical imaging tasks |

## V. RECENT ADVANCES IN LEVEL 3 OF MIC (SPECIFIC APPLICATIONS)

### A. Medical Image Data

*1) Medical imaging modalities*: Medical imaging plays a critical role in modern healthcare, offering non-invasive visualization of the human body for diagnosis and treatment planning. Various modalities, including X-ray, CT, MRI, ultrasound, PET, and SPECT, provide unique insights into different organs and tissues (Fig. 3 [70]; Fig. 4 [71] illustrates the diverse applications of these modalities across various anatomical structures.

The Medical Segmentation Decathlon dataset [72] exemplifies this diversity, encompassing 2,633 3D images spanning ten different organs (Fig. 5). Each modality possesses distinct characteristics, advantages, and limitations, necessitating careful selection based on the specific clinical scenario. Understanding these nuances is crucial for optimal utilization of medical imaging technology. A concise comparison of imaging techniques reveals their unique advantages and limitations is presented in Table XV.

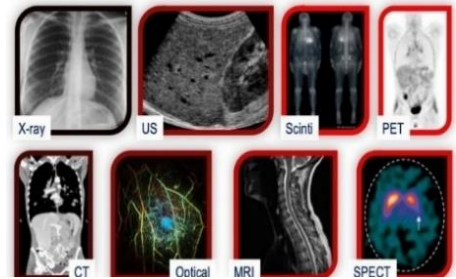

Fig. 3. Illustration of the diverse imaging techniques [70].

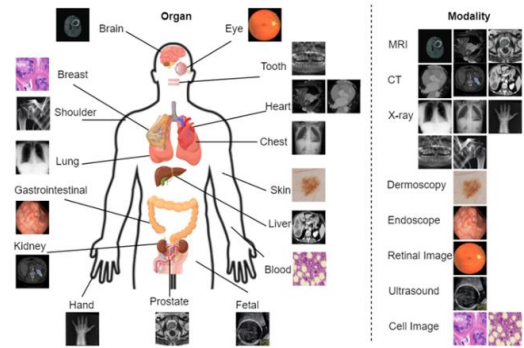

Fig. 4. An overview of the organs and corresponding medical imaging modalities [71].

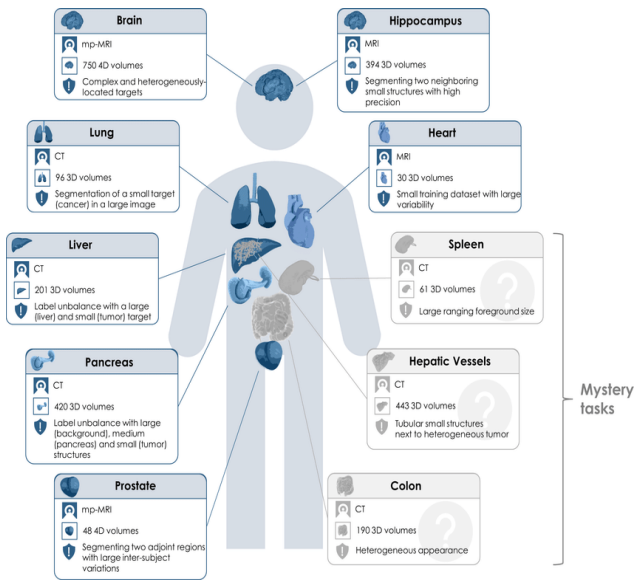

Fig. 5. An illustration of the Medical Segmentation decathlon's ten distinct tasks [72].

*2) Public databases in medical imaging research*: The growth of public medical image databases has been crucial in advancing disease classification research. Noteworthy databases include:

- ChestX-ray14[1]: Over 100,000 chest X-ray images.
- MURA[2]: More than 40,000 X-ray images of bones and joints.
- NIH Clinical Center's dataset: [3] A diverse range of modalities.
- ISIC [4]: Skin image collection for lesion detection.
- DeepLesion[5]: Nearly 10,600 CT scans.
- CheXpert:[6] Over 224,000 chest radiographs.
- MIMIC-CXR[7]: over 377,000 chest radiographs

Platforms like the World Health Data Hub of the WHO[8], Medical ImageNet[9], Kaggle[10], and PaperswithCode[11], offer extensive resources for machine learning research in medical imaging, showcasing the collaborative and open nature of contemporary scientific inquiry.

*3) Advanced techniques in medical imaging research*: Innovative computational techniques such as augmentation, transfer learning, Generative Adversarial Networks (GANs), and Federated Learning are pushing the boundaries of medical imaging research. These methods improve model performance, generate new data, and enable decentralized learning, thus enhancing the robustness and diversity of medical imaging applications.

*4) Summary*: Medical imaging is a cornerstone of modern medical diagnostics, with each modality serving specific purposes based on clinical needs. The advent of AI and machine learning, alongside the proliferation of public datasets, is revolutionizing medical imaging research, promising more accurate disease detection and personalized medicine. The future of medical imaging lies in harnessing these technological advancements to improve healthcare outcomes.

General comment

- Ionizing radiation (X-ray and CT) can be harmful, especially for pregnant women.
- MRI offers the highest detail without radiation but is expensive and not suitable for everyone.
- Ultrasound is safe and widely available but offers less detail.
- PET and SPECT provide functional information but involve radioactive materials.

*B. Advancements in Medical Imaging Diagnosis: From CAD to AI-CAD*

AI integration has profoundly transformed various domains, notably evident in medical diagnostic imaging. This shift marks a significant departure from conventional Computer-Aided Diagnosis (CAD) to AI-driven CAD systems, ushering in a new era of diagnostic capabilities,

The evolution began in the 1960s with CAD systems aiming to automate diagnostic processes. A significant milestone was the FDA's approval of a mammography CAD device by R2 Technology, Inc., in 1998, marking the start of the "CAD era." Endorsement for reimbursement by the Centers for Medicare and Medicaid Services in 2002 further accelerated CAD's adoption across modalities like chest radiographs and CT scans.

CAD systems encompass three categories based on their role in image interpretation: second-reader, concurrent-reader, and first-reader types (Fig. 6 [73]). Notably, interactive CAD falls under the first-reader type. The evolution of CAD architecture has transitioned from sequential interpretation (seen in second-reader CAD Fig. 6(a)) to simultaneous interpretation (concurrent-reader CAD Fig. 6(b)), streamlining the diagnostic process by integrating CAD results from the outset. The advent of first-reader CAD (Fig. 6(c)) presents a novel approach where CAD autonomously conducts initial interpretation, guiding the physician's analysis solely on CAD-marked images, showing promise for mass screenings like mammography.

---

[1] https://nihcc.app.box.com/v/ChestXray-NIHCC
[2] https://stanfordmlgroup.github.io/competitions/mura/
[3] https://clinicalcenter.nih.gov/
[4] https://isdis.org/
[5] https://camelyon17.grand-challenge.org/
[6] https://aimi.stanford.edu/chexpert-chest-x-rays
[7] https://physionet.org/content/mimic-cxr/2.0.0/
[8] https://www.who.int/data/
[9] https://aimi.stanford.edu/medical-imagenet
[10] https://www.kaggle.com/datasets
[11] https://paperswithcode.com/datasets

TABLE XV. COMPARISON OF MEDICAL IMAGING MODALITIES IN MIC

| Technique | Description | Pros | Cons | Safety and Image Detail |
|---|---|---|---|---|
| X-ray | Examines bones, detects fractures, tumors, and infections. | Quick, painless, cost-effective, immediate results, widely available. | Limited soft tissue contrast, ionizing radiation exposure. Not suitable for detailed organ visualization. | Radiation risk: Moderate. Image detail: Low. Best for bone visualization. |
| CT | Detailed cross-sectional images of the body. Examines organs, blood vessels, and detects abnormalities. | High-resolution images, fast acquisition time, useful for diagnosing trauma, differentiates tissue densities. | Ionizing radiation exposure. Limited soft tissue contrast compared to MRI. Not suitable for pregnant women due to radiation risks. | Radiation risk: High. Image detail: High. Excellent for visualizing organs and bone. |
| MRI | Detailed images of internal structures. Assesses brain, spinal cord, joints, and organs. | Superior soft tissue contrast, no ionizing radiation. Multiplanar imaging, detects subtle abnormalities. | Expensive, long scan times, contraindicated for patients with certain metallic implants. | Radiation risk: None. Image detail: Very High. Best for soft tissue and organ visualization. |
| Ultrasound | Uses sound waves to produce real-time images. Examines abdomen, pelvis, heart, and monitors fetal development. | Real-time imaging, non-invasive, safe, portable, widely available, no ionizing radiation. | Operator-dependent, limited penetration in obese patients, less detailed images compared to other modalities. | Radiation risk: None. Image detail: Moderate. Best for real-time imaging and pregnancy monitoring. |
| PET | Visualizes metabolic processes. Detects cancer, assesses treatment response, evaluates brain disorders. | Provides functional information, detects diseases early, helps in personalized medicine. | Expensive, limited spatial resolution, radioactive material involved. | Radiation risk: Low. Image detail: Moderate. Best for visualizing metabolic activity. |
| SPECT | Detects gamma rays emitted by a tracer. Assesses blood flow, detects myocardial infarctions, and evaluates brain disorders. | Non-invasive, provides functional information, good spatial resolution. | Longer acquisition time than PET, lower sensitivity than PET, radioactive material involved. | Radiation risk: Low. Image detail: Moderate. Best for visualizing blood flow and brain function. |

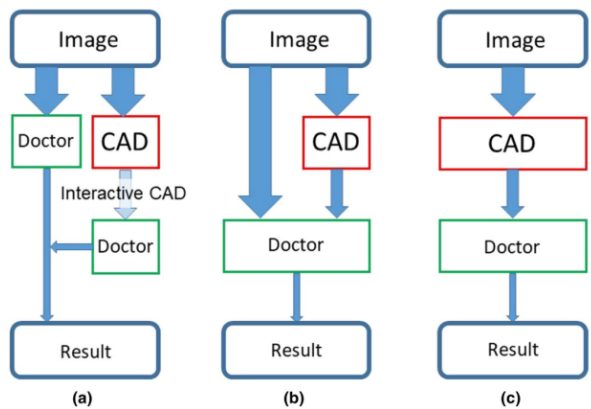

Fig. 6. Categorization of CAD systems in medical imaging interpretation: a) Second-reader, b) Concurrent-reader, and c) First-reader types [73].

Despite CAD's acknowledged utility, persistent challenges include high development costs, elevated false-positive rates leading to increased recalls and biopsies, and limited clinical efficacy. These challenges are well-documented in clinical studies, emphasizing the need for AI-driven solutions.

The recent introduction of AI-CAD, primarily employing deep learning methodologies, signifies a significant advancement. Deep learning algorithms have proven effective in reducing interpretation time and improving diagnostic accuracy, as demonstrated by studies like Kyono et al., which explored deep learning's potential to ease radiologists' workload in mammography screenings. AI-CAD's reliance on deep learning adopts a data-driven approach, benefiting from extensive datasets to enhance performance. Fig. 7 illustrates the superior performance of deep learning-based AI-CAD compared to traditional CAD systems, particularly with increasing data volume.

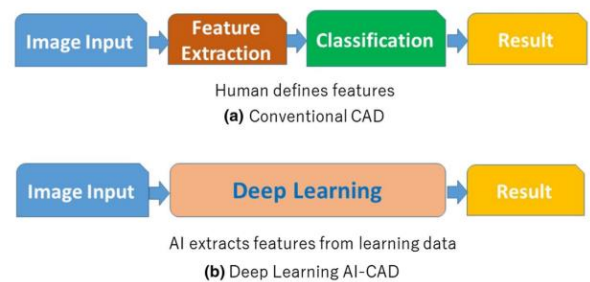

Fig. 7. Development processes: a) Conventional CAD vs. b) Deep learning-based AI-CAD [73].

In conclusion, the shift from CAD to AI-CAD represents a significant advancement in medical imaging diagnosis, offering increased accuracy, efficiency, and versatility. As AI matures, its integration has the potential to revolutionize healthcare delivery, providing clinicians with sophisticated diagnostic tools for precise and timely patient care.

Remarkable Applications of AI-CAD

- Breast Cancer: AI-CAD systems have demonstrated significant potential in breast cancer screening and mammography interpretation. Systems like cmAssist [54] can reduce false-positive markings by up to 69%, minimizing unnecessary follow-up procedures and patient anxiety. Deep learning models have shown accuracy comparable to experienced radiologists, with some hybrid models outperforming human experts. The AI-STREAM study [55] aims to generate real-world evidence on the benefits and drawbacks of AI-based computer-aided detection/diagnosis (CADe/x) for breast cancer screening.

- Tuberculosis Detection: AI-based CAD systems can assist in community-based active case finding for tuberculosis, especially in areas with limited access to experienced physicians. Okada et al. [56] demonstrated the applicability of AI-CAD for pulmonary tuberculosis in community-based active case findings, showing performance levels nearing human experts. This approach holds promise in triaging and screening tuberculosis, with significant implications for addressing healthcare professional shortages in low- and middle-income countries. Such advancements contribute to the World Health Organization's goal of "Ending tuberculosis" by 2030.

- Eye Disease Diagnosis: Google's deep learning analysis [57] achieved a detection sensitivity of about 98% in diagnosing eye diseases. AI analysis of fundus photographs [58] can assist in diagnosing not only eye diseases but also systemic conditions like heart disease, surpassing human capabilities.

- Skin Cancer Diagnosis: AI demonstrates accuracy equivalent to or higher than dermatologists in diagnosing skin cancer, utilizing deep learning on large datasets of skin lesions. Studies have shown AI achieving diagnostic accuracy comparable to dermatologists [59] and even outperforming them in differentiating melanoma [60].

- Bone diseases: The use of AI, particularly deep learning, is gaining traction in the medical community for diagnosing and treating bone diseases. Recent applications focus on segmentation and classification of bone tumors and lesions in medical images. For instance, Zhan et al. [61] developed SEAGNET, a novel framework for segmenting malignant bone tumors. Yildiz Potter et al. [62] explored a multi-task learning approach for automated bone tumor segmentation and classification. Additionally, Ye et al. [63] investigated an ensemble multi-task deep learning framework for the detection, segmentation, and classification of bone tumors and infections using multi-parametric MRI. These studies highlight the potential of deep learning to significantly improve the accuracy and efficiency of diagnosing and treating bone diseases.

- Other Pathological Applications: AI has demonstrated superior performance in detecting lymph node metastasis of breast cancer [64] and detecting diabetes from fundus photographs [65] with high sensitivity and specificity. These applications underscore AI's potential in enhancing the accuracy and efficiency of medical imaging diagnosis, ultimately improving patient outcomes and healthcare delivery.

Overall, AI-CAD systems have shown remarkable potential in various medical imaging applications, from breast cancer screening to tuberculosis detection, eye disease diagnosis, skin cancer diagnosis, and other pathological conditions. By leveraging the power of deep learning and large datasets, these systems can augment and enhance human expertise, leading to improved diagnostic accuracy, efficiency, and accessibility in healthcare.

*C. Recent Research Trends in Medical Image Classification and Cancer Statistics (2020-2024)*

Recent statistics from representative journals using keywords related to medical image classification cover the latest advancements from 2020 to 2024. In addition, the 2024 Cancer Statistics [74] indicate a 33% decrease in cancer deaths in the U.S. since 1991, attributed to reduced smoking, earlier detection, and improved treatments. However, the incidence of six major cancers continues to rise, with colorectal cancer becoming a leading cause of death among men under 50. Efforts like the Persistent Poverty Initiative aim to mitigate cancer outcomes' impact of poverty, emphasizing the need for increased investment in prevention and disparity reduction.

The document concludes with a projection of the top ten cancer types for new cases and deaths in the United States for 2024, underscoring the ongoing challenge and importance of advancements in medical imaging diagnosis.

VI. CHALLENGES AND ADVANCEMENTS IN MIC

While MIC has experienced significant progress, challenges remain in data limitations, algorithm development, and healthcare integration. This section explores these challenges and proposes innovative solutions to advance the field.

TABLE XVI. FIVE-YEAR STATISTICS OF MEDICAL IMAGE CLASSIFICATION RESEARCH IN FOUR REPRESENTATIVE JOURNALS (2020-2024)

|    | Classes | Springer | Sciencedirect | IEEE | PubMed |
|----|---------|----------|---------------|------|--------|
| 1  | cancer | 4064 | 3474 | 748 | 291 |
| 2  | brain | 3599 | 2984 | 523 | 112 |
| 3  | tumor | 2440 | 2789 | 436 | 103 |
| 4  | lesion | 2378 | 3035 | 286 | 81 |
| 5  | lung | 2374 | 2102 | 433 | 98 |
| 6  | Breast | 2019 | 1815 | 309 | 110 |
| 7  | eye | 1979 | 1602 | 144 | 39 |
| 8  | COVID | 1894 | 1460 | 343 | 107 |
| 9  | skin | 1865 | 1647 | 241 | 71 |
| 10 | Heart | 1547 | 1489 | 121 | 19 |
| 11 | AIDS | 964 | 738 | 30 | 3 |
| 12 | liver | 898 | 971 | 61 | 25 |
| 13 | bone | 847 | 938 | 83 | 32 |
| 14 | cardiac | 722 | 898 | 30 | 14 |
| 15 | prostate | 581 | 638 | 25 | 14 |
| 16 | kidney | 541 | 696 | 32 | 20 |
| 17 | tuberculosis | 471 | 321 | 49 | 4 |
| 18 | colorectal | 442 | 494 | 35 | 22 |
| 19 | Malaria | 178 | 115 | 25 | 6 |

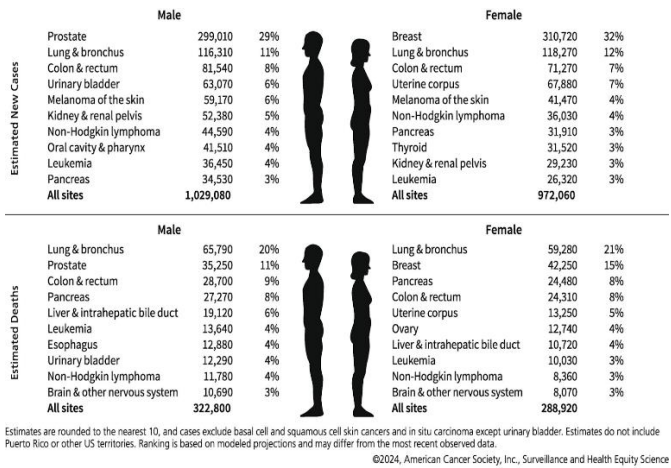

Fig. 8. Projected top ten cancer types for new cases and deaths in the United States for 2024, by gender [74].

A. *Challenges and Solutions in MIC*

  1) *Medical image data*:

  *a) Limited labeled data*: Transfer learning has shown promise in addressing the scarcity of labeled data. Kim et al. [75] provide a comprehensive review of transfer learning methods for MIC. Additionally, FSL, ZSL, and Med-VLM have been explored as potential solutions, as mentioned in previous sections. Fig. 8 shows the projected top ten cancer types for new cases and deaths in the United States.

  *b) Inter-class similarity and imbalanced datasets*: Islam et al. [76] introduced CoSSIF, a cosine similarity-based image filtering method for synthetic medical image datasets to improve accuracy when dealing with high inter-class similarity. For imbalanced datasets, Huynh et al. [77] propose a semi-supervised learning approach for MIC.

  *c) Large image sizes and domain shift*: Sreenivasulu and Varadarajan [78] present an efficient lossless ROI image compression technique to address computational challenges posed by large image dimensions. Guan and Liu [79] provide a survey on domain adaptation methods for medical image analysis, highlighting techniques to improve model generalizability across different datasets and populations.

  2) *Clinical data*:

  *a) Data privacy, security, and accessibility*: Kaissis et al. [80] discuss secure, privacy-preserving, and federated machine learning approaches in medical imaging, addressing crucial aspects of protecting patient data while improving access to clinical data.

  3) *Practical application challenges*

  *a) Model interpretability*: Alam et al. [81] explore LRP and Grad-CAM visualization techniques to interpret multi-label-multi-class pathology prediction using chest radiography, enhancing model interpretability.

  *b) Model validation*: Ramezan et al. [82] evaluate sampling and cross-validation tuning strategies for regional-scale machine learning classification, ensuring model performance and generalizability.

  *c) Regulatory approval*: Joshi and Bhandari [83] provide an updated landscape of FDA-approved AI/ML-enabled medical devices, offering insights into navigating regulatory requirements.

  Future Directions and Research Opportunities:

  This review highlights various challenges in medical image classification and presents potential solutions based on recent research. However, it's important to note that these solutions require further validation in specific clinical contexts. To advance the field, researchers should consider:

- Conducting comparative studies of different approaches to address each challenge.
- Validating the proposed solutions in diverse clinical settings and with larger datasets.
- Investigating the integration of multiple solutions to address complex, real-world scenarios in medical image classification.
- Exploring the ethical implications and potential biases of AI systems in healthcare**.**

B. *Key Advancements in MIC Techniques*

  *a) Transformers vs. CNNs*: Evidence suggests Transformers like ViT and DeiT demonstrate promising results compared to traditional CNNs, especially in capturing global context and long-range dependencies.

  *b) Synergy of transformers and CNNs*: Hybrid models like MedViT and TransMT-Net leverage the strengths of both architectures, achieving superior performance in classification and segmentation tasks.

  *c) Med-VLMs for multitask MIC*: Integrating Med-VLMs into multitask learning frameworks improves performance by effectively aligning visual and textual information.

  *d) AI for tumor classification*: AI models demonstrate impressive accuracy in distinguishing between benign and malignant tumors, with potential to augment clinical decision-making.

  *e)* FSL addresses the challenge of limited labeled data by enabling models to generalize effectively from a small number of examples. Researchs have demonstrated that FSL can achieve high accuracy in tasks such as tumor detection with minimal data, highlighting its potential in clinical applications.

  *f)* ZSL tackles the issue of classifying unseen categories by leveraging semantic relationships. ZSL has shown promising results in identifying rare diseases and novel medical conditions, significantly aiding in early diagnosis and treatment planning.

  *g)* XAI techniques enhance the interpretability and trustworthiness of MIC systems, making them more acceptable in clinical practice. Additionally, XAI contributes to optimizing model performance and accuracy by providing insights that allow for iterative model adjustments.

*C. Summary*

Addressing data challenges, refining algorithms, and ensuring responsible implementation are crucial for advancing MIC. The integration of Transformers, CNNs, Med-VLMs, and XAI techniques holds immense potential for improving healthcare delivery and patient outcomes. With the increasing focus on explainability and trustworthiness in AI models, further breakthroughs in MIC and its transformative impact on healthcare can be anticipated.

## VII. CONCLUSION AND FUTURE DIRECTIONS

The paper outlines the development of medical image classification through three solution levels: basic, specific, and applied. It discusses traditional high-performance deep learning models and highlights the promising vision-language models that can explain predictions. The paper also emphasizes the potential of multimodal models combining clinical and paraclinical data for disease diagnosis and treatment. It notes the research community's growing interest in early prediction to reduce risks and the role of Explainable Artificial Intelligence in improving predictive results. The application of AI in Computer Vision for medical purposes consistently surpasses expectations, indicating a future focus on integrating AI advancements into diagnostic and treatment-related problems using multimodal data.

ACKNOWLEDGMENT

This research is supported by Dept. of Computer Vision and Cognitive Cybernetics, Faculty of Information Technology, University of Science, Vietnam National University-Ho Chi Minh City.